\documentclass[twocolumn,pra,aps,showpacs,superscriptaddress,floatfix,showkeys,nofootinbib,10pt]{revtex4-2}

\usepackage[T1]{fontenc}
\usepackage{amsmath,amsfonts,amssymb,amsthm}
\usepackage{graphicx}

\usepackage{tikz}

\usepackage{listings}

\newcommand{\ket}[1]{ | \, #1 \rangle} \newcommand{\bra}[1]{ \langle #1 \, |} 
\newcommand{\proj}[1]{\ket{#1}\bra{#1}}

\newcommand{\Ab}[1]{ \left| #1 \, \right|} 
\newcommand{\be}{\begin{equation}} \newcommand{\ee}{\end{equation}}
\newcommand{\ba}{\begin{aligned}} \newcommand{\ea}{\end{aligned}}

\newcommand{\Max}[1]{\mathrm{Max}[#1]}
\newcommand{\Min}[1]{\mathrm{Min}[#1]}
\newcommand{\Var}[1]{\mathrm{Var}[#1]}

\DeclareMathOperator{\Tr}{Tr}

\DeclareRobustCommand\openone{\leavevmode\hbox{\small1\normalsize\kern-.33em1}}%

\newtheorem{dfn}{Definition}
\newtheorem{trm}{Theorem}

\usepackage{color}
\definecolor{mygreen}{rgb}{0,0.6,0}
\definecolor{mylightgray}{rgb}{0.98,0.98,0.98}
\definecolor{mygray}{rgb}{0.5,0.5,0.5}
\definecolor{mymauve}{rgb}{0.58,0,0.82}
\begin{document}

\title{Finite-size security analysis for quantum protocols: A Python framework using the Entropy Accumulation Theorem with graphical interface}

\author{P.~Mironowicz} \email{piotr.mironowicz@gmail.com}
\affiliation{Center for Theoretical Physics, Polish Academy of Sciences, Aleja Lotników 32/46, 02-668 Warsaw, Poland}
\affiliation{Department of Algorithms and System Modeling, Faculty of Electronics, Telecommunications and Informatics, Gda\'{n}sk University of Technology, Poland}
\affiliation{Department of Physics, Stockholm University, S-10691 Stockholm, Sweden}

\author{Mohamed Bourennane}
\affiliation{Department of Physics, Stockholm University, S-10691 Stockholm, Sweden}

\date{\today}

\begin{abstract}
We present a comprehensive software framework for the finite-size security analysis of quantum random number generation (QRNG) and quantum key distribution (QKD) protocols, based on the Entropy Accumulation Theorem (EAT). Our framework includes both a Python API and an intuitive graphical user interface (GUI), designed to support protocol designers and experimentalists in certifying randomness and key rates under realistic, finite-resource conditions. At its core, the framework automates the construction of min-tradeoff functions via semi-definite programming and integrates them into a full entropy analysis pipeline. Users can specify device configurations, Bell-type inequalities or probability constraints, and select entropy measures such as min-entropy or von Neumann entropy. The package further provides tools for setting test parameters, computing secure randomness rates, and exploring tradeoffs between statistical confidence, protocol duration, and randomness output. We demonstrate the framework showing how users can move from theoretical constraints to practical security bounds with minimal overhead. This work contributes a reproducible, modular, and extensible platform for certifying quantum protocols under finite-size effects, significantly lowering the skill barrier to rigorous quantum cryptographic analysis.
\end{abstract}


\maketitle

\section{Introduction}\label{sec:introduction}

Randomness underpins the security of modern cryptography, from key generation to secure communication. Yet classical sources of randomness are often predictable if an adversary gains partial knowledge of the system. Quantum processes, by contrast, naturally exhibit indeterministic behavior, making them ideal candidates for generating truly unpredictable, private random numbers. In particular, quantum random number generation (QRNG) exploit fundamental quantum events to produce randomness, while quantum key distribution (QKD) uses quantum correlations—such as those witnessed in entanglement—to establish secret keys between distant parties with theoretically provable security~\cite{gisin2002quantum,scarani2009security,nielsen2010quantum,pirandola2020advances}.

Device-independent (DI) protocols enhance security by not assuming trust in the internal workings of quantum devices: instead, violations of Bell inequalities certify both randomness and privacy, even in the presence of untrusted hardware~\cite{pironio2010random,acin2016certified}. However, practical implementations invariably operate in finite-runtime conditions without independent and identically distributed (i.i.d.) assumptions~\cite{tomamichel2016quantum}. Addressing these challenges requires applying rigorous frameworks like the Entropy Accumulation Theorem (EAT) or Quantum Probability Estimation (QPE)~\cite{dupuis2020entropy,knill2018quantum}.

Despite its strength, the EAT remains challenging to learn and apply correctly. The underlying mathematics—particularly involving min-tradeoff functions, randomness bounds, and error terms—can be daunting for many practitioners. Additionally, for experimentalists to develop custom implementations of EAT applications and their associated interfaces from scratch is a demanding task in terms of time and resources. To bridge this gap, we introduce a comprehensive Python-based toolkit—complete with both application programming interface (API) and graphical user interface (GUI)—implementing EAT-based analysis for QRNG and QKD. This framework not only automates complex calculations, but also integrates configurable parameters, real-time plotting, and result export features, significantly lowering the barrier for experimental deployment.

The remainder of this paper is structured as follows. In Section~\ref{sec:eat}, we provide an overview of the EAT, beginning with a historical and conceptual summary in Subsection~\ref{subsec:eat-development}, followed by a discussion of the core parameters relevant to its application in Subsection~\ref{sec:eat_parameters}. We then explain the notion of EAT channels and the Markov condition in Subsection~\ref{sec:eat_channels}, introduce the concept of min-tradeoff functions in Subsection~\ref{sec:min_tradeoff}, and finally present a formal statement of the theorem in Subsection~\ref{sec:eat_statement}.

Section~\ref{sec:software} describes the design and key functionalities of our software platform developed to facilitate EAT-based security analyses. This is followed by two illustrative case studies. In Section~\ref{sec:case-study}, we explore a QRNG protocol based on the modified CHSH scenario, using min-entropy and the API for automation. Section~\ref{sec:gui} then examines a QKD protocol in the standard Clauser–Horne–Shimony–Holt (CHSH) scenario~\cite{clauser1969proposed}, using von~Neumann entropy and demonstrating how to conduct a full EAT-based analysis via GUI.
Section~\ref{sec:conclusions} concludes the paper with a summary and future directions.


\section{Entropy Accumulation Theorem overview}\label{sec:eat}

In DI protocols for QRNG or QKD, one interacts sequentially with untrusted devices over \(n\) rounds. The EAT provides a way to bound the total smooth min-entropy accumulated across these rounds, based on per-round statistics and a suitable ``min-tradeoff'' function. We follow the notation of Brown, Ragy, and Colbeck~\cite{brown2019framework}, which matches the formulas used in our code.

\subsection{Development of the Entropy Accumulation Theorem}\label{subsec:eat-development}

The EAT was first established by Dupuis, Fawzi, and Renner in a rigorous mathematical framework, providing a way to bound the total smooth min-entropy in sequential quantum processes~\cite{dupuis2020entropy}. Subsequently, Dupuis and Fawzi improved the finite-size analysis by deriving tighter second-order correction terms, refining the error bounds in practical settings~\cite{dupuis2019entropy}. More recently, Metger, Fawzi, Sutter, and Renner extended EAT to a more general setting, allowing broader classes of channels and side-information models in their generalized entropy accumulation result~\cite{metger2024generalised}. In parallel, Arqand, Hahn, and Tan introduced a generalized Rényi entropy accumulation theorem together with a unified quantum probability estimation framework, further broadening the applicability of entropy accumulation techniques~\cite{arqand2024generalized}. On the experimental and applied front, Arnon-Friedman, Dupuis, Fawzi,~\emph{et al.} demonstrated how EAT enables practical device-independent quantum cryptography, showing its feasibility under realistic noise and finite-size constraints~\cite{arnon2018practical}. More recently, Metger and Renner applied generalized entropy accumulation to prove security of QKD under minimal assumptions, highlighting EAT’s central role in advancing QKD proofs~\cite{metger2023security}. Arnon-Friedman, Renner, and Vidick provided security proofs leveraging entropy accumulation, streamlining the conceptual and technical requirements for DI protocols~\cite{arnon2019simple}. Brown, Ragy, and Colbeck formulated a comprehensive framework for device-independent randomness expansion using EAT, codifying how to derive min-tradeoff functions and optimize parameters in practice~\cite{brown2019framework}.

Several experimental and application-focused works have since utilized these EAT-based methods. For example, Liu~\emph{et al.} implemented DI randomness expansion against quantum side information in a physical setup, demonstrating high-rate certified randomness via EAT bounds~\cite{liu2021device}. Likewise, Nadlinger~\emph{et al.} reported an experimental QKD protocol certified by Bell’s theorem, where EAT-based finite-size analysis underpinned the security guarantees~\cite{nadlinger2022experimental}. Together, these developments illustrate the evolution and broad applicability of entropy accumulation in quantum cryptography and randomness generation.

\subsection{Parameters of EAT}\label{sec:eat_parameters}

A key parameter in EAT is $\epsilon_{\text{s}}$, commonly referred to as the smoothing parameter of randomness. Smaller values of $\epsilon_{\text{s}}$ correspond to higher-quality certified randomness. Another central parameter is the observed Bell value, which quantifies the degree of violation of a chosen Bell inequality~\cite{brunner2014bell}. Higher Bell values typically signal improved experimental quality.

The efficiency of the system is captured by the \textit{events per second} parameter, representing the number of entangled pairs successfully detected per second. Since practical quantum protocols operate under time constraints, EAT accounts for the impact of finite data by introducing the \textit{single data chunk generation time}, which specifies the duration over which data is collected for a single evaluation cycle.

In spot-checking protocols, a fraction $\gamma$ of rounds are designated as test rounds, where inputs $x$ and $y$, of Alice and Bob, respectively, are chosen uniformly at random to estimate the Bell violation~\cite{miller2017universal}. While increasing $\gamma$ improves the statistical reliability of this estimate, it also consumes more randomness itself and reduces the number of rounds available for generating new private randomness. The parameter $p_\Omega$ denotes the completeness level of the protocol, i.e., the probability that the protocol does not abort, factoring in the statistical nature of Bell value estimation. The interplay between $\gamma$ and $p_\Omega$ significantly influences the achievable randomness generation rate and must be carefully balanced.

In experimental implementations, reconfiguring the optical components between test and generation rounds often introduces delays, thereby reducing the effective event rate. This overhead is captured by the \textit{switch delay}, which accounts for the slowdown induced by setting changes. Finally, EAT allows the optimization of a freely tunable parameter, denoted $\beta$, which influences the tightness of the resulting entropy bounds. A detailed discussion of the impact of these parameters on the performance of DI protocols is discussed in~\cite{piveteau2024optimization}.

\subsection{EAT channels and Markov condition}\label{sec:eat_channels}

\begin{dfn}[EAT channels]\label{def:eat-channels}
A set of \emph{EAT channels} \(\{\mathcal{N}_i\}_{i=1}^n\) is a collection of completely-positive trace-preserving maps
\begin{equation}
	\mathcal{N}_i: \mathcal{S}(R_{i-1}) \;\longrightarrow\; \mathcal{S}(A_i B_i X_i Y_i C_i R_i),
\end{equation}
where for each round \(i\in[n]\):
\begin{enumerate}
	\item \(A_i,B_i,X_i,Y_i,C_i\) are finite-dimensional \emph{classical} registers, and \(C_i\) is the output of a deterministic function of the classical registers \((A_i,B_i,X_i,Y_i)\);
	\item \(R_i\) is an arbitrary quantum register;
	\item For any initial state \(\rho_{R_0 E}\) (with \(E\) the adversary register), the final global state
		\begin{equation}
			\rho_{A^n B^n X^n Y^n C^n E} := \Tr_{R_n} \Bigl[ \bigl(\mathcal{N}_n \circ \cdots \circ \mathcal{N}_1 \otimes \mathrm{id}_E\bigr)(\rho_{R_0 E})\Bigr]
		\end{equation}
		satisfies, for every \(i\in[n]\), the Markov chain condition
		\begin{equation}
			\label{eq:MarkovChain}
			I\bigl(A^{i-1}B^{i-1} : X_i Y_i \mid X^{i-1}Y^{i-1} E\bigr) = 0.
		\end{equation}
\end{enumerate}
\end{dfn}

The Markov chain condition~\eqref{eq:MarkovChain} implies that the choice of inputs \((X_i,Y_i)\) at round \(i\) is (conditionally) independent of all previous outputs \((A^{i-1},B^{i-1})\), given past inputs \((X^{i-1},Y^{i-1})\) and the side information \(E\). This models the honest parties choosing their inputs “freshly” each round, without leaking through the devices.

\subsection{Min-tradeoff functions}\label{sec:min_tradeoff}

To apply EAT, one needs a \emph{min-tradeoff function} that lower-bounds the per-round conditional von Neumann entropy in terms of the observed classical outputs \(C_i\). Roughly, \(C_i\) encodes one round’s statistics (e.g.a Bell-violation score).

\begin{dfn}[Min-tradeoff function]\label{def:fmin}
	Let \(\{\mathcal{N}_i\}_{i=1}^n\) be EAT channels with common classical-alphabet register \(C_i\in\mathcal{C}\), and \(\mathcal{P}_{\mathcal{C}}\) be the set of all probability distributions over \(\mathcal{C}\). An affine function
	\begin{equation}
		f: \mathcal{P}_{\mathcal{C}} \;\to\; \mathbb{R}
	\end{equation}
	is a \emph{min-tradeoff function} if for each round \(i\) and for any state \(\sigma_{R_{i-1}R'}\) (with \(R'\simeq R_{i-1}\)) such that the marginal on \(C_i\) after applying \(\mathcal{N}_i\) is the classical distribution \(\tau_p = \sum_{c\in\mathcal{C}} p(c)\proj{c}\), one has
	\begin{equation}
		f(p) \;\le\; \inf_{\sigma_{R_{i-1}R'}:\,\mathcal{N}_i(\sigma)_{C_i} = \tau_p}
		 H\bigl(A_i B_i \,\big|\, X_i Y_i R'\bigr)_{\mathcal{N}_i(\sigma)}.
	\end{equation}
	If no state yields \(\mathcal{N}_i(\sigma)_{C_i}=\tau_p\), the infimum is taken as \(+\infty\). 
\end{dfn}

In practice, one constructs \(f\) by solving suitable SDP, e.g.using Navascu{\'e}s-Pironio-Ac{\'\i}n (NPA) or BFF relaxations, that lower-bound the conditional entropy per round given observed statistics~\cite{navascues2007bounding,navascues2008convergent,mironowicz2024semi}.

\subsection{EAT statement}\label{sec:eat_statement}

Let \(\rho_{A^n B^n X^n Y^n C^n E}\) be the overall state after applying the EAT channels sequentially to an initial \(\rho_{R_0 E}\). Suppose the parties observe classical outputs \(C^n = (C_1,\dots,C_n)\) and consider an event \(\Omega \subseteq \mathcal{C}^n\) (e.g.\ that the empirical frequency of outcomes yields a sufficient Bell-violation score). Write \(p_\Omega = \Pr[C^n \in \Omega]\) and let \(\rho_{|_\Omega}\) be the post-selected state conditioned on \(\Omega\).

\begin{trm}[Entropy Accumulation Theorem~\cite{dupuis2019entropy,brown2019framework}]
\label{thm:EAT}
Let \(f\) be a valid min-tradeoff function for the channels \(\{\mathcal{N}_i\}\). Suppose that for every observed sequence \(C^n\in\Omega\) (with nonzero probability), the empirical frequency \(\mathrm{freq}(C^n)\) satisfies 
\begin{equation}
	f\bigl(\mathrm{freq}(C^n)\bigr) \ge t
\end{equation}
for some real \(t\). Then for any smoothing parameter \(\epsilon_{\text{s}}\in(0,1)\) and any \(\beta\in(0,1)\), the smooth min-entropy of the outputs \(A^n B^n\) conditioned on inputs \(X^n Y^n\) and side information \(E\) satisfies
\begin{equation}\label{eq:EAT-bound}
	H_{\min}^{\epsilon_{\text{s}}}(A^n B^n \mid X^n Y^n E)_{\rho_{|_\Omega}}
	\;>\; n\,t \;-\; n\,(\epsilon_V + \epsilon_K) \;-\; \epsilon_\Omega,
\end{equation}
where the correction terms are:
\begin{subequations}
	\label{eqs:eat_epses}
	\begin{equation}
		\epsilon_V := \frac{\beta \ln 2}{2}\Bigl(\log\bigl(2|\mathcal{A}\mathcal{B}|^2 + 1\bigr) + \sqrt{\Var{\left.f\right|_{\Gamma}} + 2}\Bigr)^2,
	\end{equation}
	\begin{equation}
		\epsilon_K := \theta_1 \times \theta_2 \times \theta_3,
	\end{equation}
	\begin{equation}
		\epsilon_\Omega := \frac{1}{\beta}\bigl(1 - 2\log(p_{\Omega}\,\epsilon_{\text{s}})\bigr),
	\end{equation}
\end{subequations}
where
\begin{subequations}
	\begin{equation}
		\theta_1 := \frac{\beta^2}{6(1-\beta)^3 \ln 2},
	\end{equation}
	\begin{equation}
		\theta_2 := 2^{\beta\bigl(\log|\mathcal{A}\mathcal{B}| + d_f \bigr)},
	\end{equation}
	\begin{equation}
		\theta_3 := \ln^3\Bigl(2^{\log|\mathcal{A}\mathcal{B}| + d_f} + e^2\Bigr),
	\end{equation}
	\begin{equation}
		\label{eq:diameter_f}
		d_f := \Max f - \Min{\left.f\right|_{\Gamma}}.
	\end{equation}
\end{subequations}
Here \(\Ab{\mathcal{A}\mathcal{B}}\) denotes the size of the alphabet of \((A_i,B_i)\) per round.
\(\Var{\left.f\right|_{\Gamma}}\) is the worst-case variance of \(f(C_i)\) over the conditional distribution induced by \(\Omega\).
\(\Max f\) and \(\Min{\left.f\right|_{\Gamma}}\) are the maximum and minimum of \(f\) over the relevant probability simplex.
\end{trm}

The bound \eqref{eq:EAT-bound} shows that, conditioned on the observed event \(\Omega\), one accumulates at least \(n t\) bits of min-entropy up to finite-size penalties \(\epsilon_V\), \(\epsilon_K\), and \(\epsilon_\Omega\). One chooses \(\beta\) to optimize the trade-off between the variance term \(\epsilon_V\), the higher-order term \(\epsilon_K\), and the smoothing penalty \(\epsilon_\Omega\).

For precise definitions of \(\Var{\left.f\right|_{\Gamma}}\), \(\Max f\), \(\Min{\left.f\right|_{\Gamma}}\), the Markov condition, and technical proofs, see~\cite{dupuis2019entropy,brown2019framework}.  
In practice, one defines the EAT channels based on the protocol’s measurements and chooses \(C_i\) to encode the Bell-score or other relevant statistics.  
The min-tradeoff function is giving a linear (or affine) lower bound on \(H(A_iB_i\mid X_iY_i E)\) as a function of \(C_i\)’s distribution.  
The final extracted randomness (or key rate) per round follows from the bound \eqref{eq:EAT-bound}, subtracting any additional costs, e.g.\ error correction leakage in QKD.

This brief overview sets up the notation and formulas used in the \texttt{expdiqrng} package for finite-size analysis via EAT, following the framework of Brown~\emph{et al.}~\cite{brown2019framework}.

\section{Software design and functionality}\label{sec:software}

The \texttt{expdiqrng} Python package is structured into two main subpackages: \texttt{expdiqrng} and \texttt{expdiqrng\_ui}. This section describes the architecture and functionality of the \texttt{expdiqrng} subpackage, which provides the core numerical and theoretical tools for analyzing experimental data in the context of device-independent quantum randomness generation.

The \texttt{expdiqrng} subpackage implements various models and computational methods from quantum information theory. It includes the following modules and associated classes:

\begin{itemize}
    \item \texttt{Eber\_data}: A data model class representing experimental outcomes, providing methods for loading, processing, and interpreting raw click data. It works in tandem with the \texttt{PostProcessingMethod} class to facilitate various post-processing strategies.
    
    \item \texttt{NetGainCalculator}: A class to evaluate the net gain rate in randomness expansion protocols, incorporating correction terms, entropy rates, and visualizations. In our code, the terms \texttt{epsV}, \texttt{epsK}, and \texttt{epsOmega} correspond to the functions~\eqref{eqs:eat_epses}.
    
    \item \texttt{MinTradeoffFunction}: Implements methods required for entropy accumulation via the EAT. It computes the tangent surface and Tsirelson bounds necessary for entropy estimation.
    
    \item \texttt{BFF (Brown-Fawzi-Fawzi)}: Implements a dualization method for estimating von Neumann entropy in quantum systems. It depends on \texttt{MinTradeoffFunction} and is used for entropy certification from observed data. The BFF method was introduced in~\cite{brown2024device} and is briefly discussed in Appendix~\ref{sec:BFF}.
    
    \item \texttt{NietoSilleras}: Uses the Navascués-Pironio-Acín (NPA) method to perform randomness certification, including the estimation of guessing probabilities and entropy bounds in the Nieto-Silleras variant~\cite{nieto2014using,bancal2014more}.
    
    \item \texttt{NS\_probabilities}: Represents non-signaling probability distributions and supports entropy calculation, uniform distribution generation, and probability expression evaluation. It depends on both \texttt{Eber\_data} and \texttt{NS\_ProbabilityMarginals}.
    
    \item \texttt{NS\_ProbabilityMarginals}: Handles marginal probability distributions, ensuring compliance with the no-signaling principle. It offers functionality to compute marginals and correlators.
    
    \item \texttt{utils.py}: A utility module containing helper functions for optimization, entropy calculations, linear algebra operations, and probability expression management. In particular, it contains the function \texttt{calculate\_probability\_expression\_val\_error} responsible for calculation error bars of estimation of the values of Bell expressions for given confidence intervals, using the method described in~\cite{pironio2010random}.
\end{itemize}

Each class is designed to encapsulate specific aspects of DI QRNG and QKD. The design prioritizes modularity, making it easy to test individual components or replace parts of the computational pipeline.

Notably, several classes such as \texttt{BFF} and \texttt{NietoSilleras} build upon the foundational \texttt{MinTradeoffFunction}, demonstrating an inheritance-like compositional design. The use of stateless helper functions in \texttt{utils.py} ensures reusability and separation of concerns.

NPA and BFF used the package ncpol2sdpa~\cite{wittek2015algorithm,Ncpol2sdpaGitHub}. The min-tradeoff functions were calculated using the dual SDP solutions, as described in Section~3.4 of~\cite{mironowicz2024semi}.

To install the package download the source, browse to its directory and run \texttt{python -m pip install .}. The code was developed using Python~3.12.

\section{Case study: QRNG in modCHSH scenario for min-entropy with API}\label{sec:case-study}

This case study demonstrates the application of the \texttt{expdiqrng} package in a DI-QRNG protocol using the so-called modCHSH Bell expression introduced in~\cite{mironowicz2013robustness} and given by:
\begin{equation}
	\label{eq:modCHSH}
	\text{modCHSH} = C(0,0) + C(0,1) + C(1,0) - C(1,1) + C(2,1).
\end{equation}
Here, \(C(x, y)\) denotes the correlator between Alice’s and Bob’s outputs for inputs \(x\) and \(y\), respectively. This expression captures a linear combination of correlators tailored to the spot-checking randomness generation scenario. We use the min-entropy as the security quantifier and evaluate finite-size effects via the EAT.

The following code snippets implement a complete workflow for computing a min-tradeoff function and subsequently estimating secure randomness rates using the EAT. Each step in the pipeline contributes to a layered process, beginning with the specification of a quantum experimental configuration and culminating in the derivation of certified randomness rates under finite-size constraints. Start with importing needed functions, viz. \texttt{calculate\_mintradeoff} and \texttt{MinTradeoffInfo} from the module \texttt{expdiqrng\_ui}, as shown in Listing~\ref{python:edq_example1_step0}.

\begin{lstlisting}[caption={QRNG example: Import expdiqrng module.},label={python:edq_example1_step0},language=Python,float,firstnumber=1]
from typing import Dict, List, Tuple # Python typing package (to increase the readability of types)
from expdiqrng_ui import calculate_mintradeoff, MinTradeoffInfo # the expdiqrng package
\end{lstlisting}

\subsection{Quantum Setup and Certificate Description}

\begin{lstlisting}[caption={QRNG example: Describe quantum setup and certificate.},label={python:edq_example1_step1},language=Python,float,firstnumber=1]
A_config: List[int] = [2, 2, 2] # Alice has 3 settings, binary outcomes
B_config: List[int] = [2, 2] # Bob has 2 settings, binary outcomes
certificate_str: str = "C(0,0) + C(0,1) + C(1,0) - C(1,1) + C(2,1)" # can also use e.g. "P(a,b|x,y)", or "PA(a|x)"
\end{lstlisting}

To begin with, the code in Listing~\ref{python:edq_example1_step1} defines the measurement settings for two parties — Alice and Bob — involved in a quantum randomness generation protocol. Specifically, Alice is configured to choose between three possible measurement settings, each yielding binary outcomes, which is reflected in the list \texttt{A\_config = [2, 2, 2]}. In parallel, Bob is limited to two binary settings, specified by \texttt{B\_config = [2, 2]}.

Simultaneously, the Bell-type certificate used for randomness certification is described using a string representation of a correlation expression. Here, the expression \texttt{"C(0,0) + C(0,1) + C(1,0) - C(1,1) + C(2,1)"} refers to a specific linear combination of correlators — known as the modCHSH inequality~\eqref{eq:modCHSH} — that serves as the statistical witness for nonlocal correlations. We note that a convenient notation of typing Bell expressions as strings is provided.

\subsection{Protocol Parameters}

\begin{lstlisting}[caption={QRNG example: Describe parameters of the protocol.},label={python:edq_example1_step2},language=Python,float,firstnumber=1]
spot_setting = (2, 0)
relaxation_level: int = 2 # level of NPA
m_radau: int = 0 # for BFF hierarchy, otherwise ignored
additional_data_dict: dict = {} # place metadata here
entropy_type_str: str = "min-entropy" # or "von Neumann entropy"
use_case_str: str = "Randomness Generation"# or "Key Distribution"
hab_dict: Dict[Tuple[int,int], float] = {spot_setting : 0.02} # values of entropy H(A|B), needed for error correction in QKD
setup_nickname: str = "modCHSH example" # any name
\end{lstlisting}

Next, as shown in Listing~\ref{python:edq_example1_step2}, the protocol-specific parameters are defined to contextualize the randomness generation scenario. Among these, the variable \texttt{spot\_setting = (2, 0)} selects the pair of measurement settings used in generation rounds, which is crucial in spot-checking protocols.

Concurrently, the level of the NPA hierarchy used to approximate the quantum set is chosen via \texttt{relaxation\_level = 2}, with an optional BFF hierarchy parameter \texttt{m\_radau}, here set to zero. Additionally, the dictionary \texttt{hab\_dict} specifies the conditional entropy \(H(A|B)\) for the chosen spot setting, which becomes particularly relevant when analyzing error correction in QKD, but for QRNG is irrelevant, shown here only for the sake of completeness of the discussion. The entropy measure of interest is declared as \texttt{"min-entropy"}, and the intended application is identified as \texttt{"Randomness Generation"} rather than the other option, viz.~\texttt{"Key Distribution"}. The string \texttt{setup\_nickname} is used purely for labeling this configuration for the user.

\subsection{Tradeoff Function Computation}

\begin{lstlisting}[caption={QRNG example: Create min-tradeoff function (from dual SDP).},label={python:edq_example1_step3},language=Python,float,firstnumber=1]
min_tradeoff_info_str, _ = calculate_mintradeoff(  # calculate_mintradeoff returns a serialized to str instance of MinTradeoffInfo
    probability_expression_str_list = [certificate_str],
    probability_expression_val_list = [3.8],
    A_config = A_config, B_config = B_config, spot_setting = spot_setting,
    relaxation_level = relaxation_level, m_radau = m_radau, additional_data_dict = additional_data_dict,
    entropy_type_str = entropy_type_str, use_case_str = use_case_str,
    hab_dict = hab_dict, setup_nickname = setup_nickname)
min_tradeoff_info = MinTradeoffInfo.from_str(min_tradeoff_info_str) # deserialize str to MinTradeoffInfo
\end{lstlisting}

Following the parameter setup, the min-tradeoff function is computed, as shown in Listing~\ref{python:edq_example1_step3}. This function plays a central role in quantifying entropy in finite-size scenarios using EAT. The method \texttt{calculate\_mintradeoff()} receives all the necessary inputs — including the certificate value (here fixed at 3.8 for illustration) — and returns a serialized version of a \texttt{MinTradeoffInfo} object. This object is then deserialized using \texttt{MinTradeoffInfo.from\_str()}, yielding a usable object that encapsulates the min-tradeoff function for the certifiable entropy.

\subsection{EAT Parameters Setup}

\begin{lstlisting}[caption={QRNG example: Set EAT parameters.},label={python:edq_example1_step4},language=Python,float,firstnumber=1]
single_data_chunk_generation_time_list: List[float] = [3600] # 1 hour time of collecting data for randomness generation
events_per_second_list: List[float] = [1e6] # how many events used in QRNG occurr per second
epsS_list: List[float] = [1e-12] # smoothing parameter of the certified entropy
confidence_interval_list: List[float] = [0.99] # list of considered p_Omega parameters, 1-p_Omega is the completeness error
test_round_probability_list: List[float] = [0.01] # list of considered gamma parameters, the propability of test round in spot-checking protocol
\end{lstlisting}

At this stage, attention shifts toward configuring the statistical parameters required by the EAT framework, as shown in Listing~\ref{python:edq_example1_step4}. These include the data acquisition time per chunk (set to 3600 seconds, i.e., one hour), and the event detection rate (set to \(10^6\) events per second). 

The smoothing parameter $\epsilon_{\text{s}} = 10^{-12}$ is used, by placing it on the list\texttt{epsS\_list = [1e-12]}. Next, the confidence level, \texttt{confidence\_interval\_list = [0.99]}, specifies that the protocol accepts a completeness error of 1\%. The value is used in this example but can be adjusted as needed. Similarly, the test round probability \texttt{[0.01]} — representing the chance that a given round is used for testing rather than generation — is also an example value that may be tuned according to experimental design. Since most of the parameters are provided as lists, if a list contain more then one element, then the EAT calculations are performed for all possible combination of the provided values.

\subsection{Finite-Size EAT Analysis}

\begin{lstlisting}[caption={QRNG example: Finite size analysis with EAT.},label={python:edq_example1_step5},language=Python,float,firstnumber=1]
rate_calculation_results, _, _ = min_tradeoff_info.calculate_eat_rates( # rate_calculation_results stores EAT results for all input parameters
    single_data_chunk_generation_time_list = single_data_chunk_generation_time_list,
    events_per_second_list = events_per_second_list,
    epsS_list = epsS_list,
    confidence_interval_list = confidence_interval_list,
    test_round_probability_list = test_round_probability_list)
net_gain_per_second, net_gain_per_second_parameters_values_dict = rate_calculation_results.get_just_net_gain_per_second() # get optimal parameters for EAT
\end{lstlisting}

With both the min-tradeoff function and EAT parameters in place, the code proceeds to evaluate the secure randomness generation rate while accounting for finite-size effects, as shown in Listing~\ref{python:edq_example1_step5}. The method \texttt{calculate\_eat\_rates()} integrates all previously defined quantities to determine how much randomness can be securely extracted under the given assumptions.

The resulting object stores detailed EAT results, from which the code extracts the net gain of randomness per second. This value reflects the protocol’s actual performance in practice. Additionally, the specific parameter configuration that achieves this optimal result is returned as a dictionary, allowing for reproducibility and fine-tuning.

\subsection{Output and Result Interpretation}

\begin{lstlisting}[caption={QRNG example: Print results.},label={python:edq_example1_step6},language=Python,float,firstnumber=1]
print(min_tradeoff_info.asymptotic_keyrate) # asymptotic rate, i.e. without finite-size effects
print(net_gain_per_second) # randomness obtained with optimal parameters
print(net_gain_per_second_parameters_values_dict) # parameters of EAT giving the optimal randomness
\end{lstlisting}

Finally, as shown in Listing~\ref{python:edq_example1_step6}, the script prints out the main outputs of the analysis. The first is the asymptotic keyrate (\texttt{min\_tradeoff\_info.asymptotic\_keyrate}), which represents the idealized rate assuming an infinite number of rounds and no statistical error. This is followed by the secure net gain per second (\texttt{net\_gain\_per\_second}), which is the practically attainable rate under the given finite-size constraints.

Lastly, the script reports the dictionary \texttt{net\_gain\_per\_second\_parameters\_values\_dict}, which holds the exact input parameters that maximize the net gain, thereby offering insight into the optimal operation point of the protocol.

The output of the API for the modCHSH scenario is shown in Listing~\ref{listing:modCHSH_output}.

\begin{lstlisting}[language=Python, caption={API output for modCHSH scenario}, label={listing:modCHSH_output}]
1.4368663908110753
947239.7510144893
{
  'diameter_of_min_tradeoff': 66.26207295685808,
  'pxpy_randomness_consumption_per_round': 2.584962500721156,
  'hab': None,
  'subtract_consumption_for_test_rounds': False,
  'min-tradeoff certificate value': 1.4368663908110753,
  'epsS': 1e-12,
  'events per second': 1000000.0,
  'single data chunk generation time': 3600.0,
  'pOmega': 0.99,
  '-log beta': 21.0,
  'test round probability': 0.01,
  'switch delay': 0.0,
  'entropy_lower_bound_const_values': 1.088404255614198
}
\end{lstlisting}

The key value of interest is the \emph{asymptotic generation rate per event}, which is approximately \(1.4369\). Given that the system operates at \(10^6\) events per second, this corresponds to an \emph{asymptotic maximal randomness generation rate} of about \(1{,}436{,}866\) bits per second.
However, due to \emph{finite-size effects} accounted for by the EAT, the \emph{actual secure randomness generation rate} is reduced to approximately \(947{,}240\) bits per second. This demonstrates the significance of EAT corrections in practical scenarios.

The entry \texttt{diameter\_of\_min\_tradeoff} indicates an upper bound on the value \(d_f\) in~\eqref{eq:diameter_f}, and has a value of approximately \(66.26\). This parameter contributes to the finite-size correction term in the entropy bound.
The field \texttt{pxpy\_randomness\_consumption\_per\_round} specifies that about \(2.585\) bits of randomness are consumed per round to randomize measurement settings and determine whether a round is for testing or generation. Since \texttt{subtract\_consumption\_for\_test\_rounds} is \texttt{False}, this consumption is not subtracted from the generated randomness, meaning that the output reflects the \emph{gross randomness gain}—assuming the input randomness is available for free.
The optimal EAT parameter is \(\beta = 2^{-21} \approx 4.768 \times 10^{-7}\).
Finally, \texttt{entropy\_lower\_bound\_const\_values} gives the constant term of the min-tradeoff function, approximately \(1.0884\), representing a contribution to the entropy lower bound that does not depend on the observed frequencies.

\section{Case study: QKD in the CHSH Scenario with von~Neumann Entropy and Graphical Interface}\label{sec:gui}

\begin{figure}
    \centering
    \includegraphics[width=\linewidth]{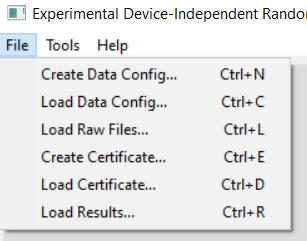}
    \caption{Menu with EAT analysis options in execution order, including keyboard shortcuts.}
    \label{fig:menu}
\end{figure}

This section presents a practical case study of applying the EAT in the context of QKD using the CHSH scenario~\cite{acin2006efficient}. We demonstrate how to use GUI to carry out the analysis step-by-step, from configuring data inputs to computing key generation rates. In Fig.~\ref{fig:menu} we see the "File" menu, where you can see options related to analysis with EAT. They are placed in the order of actions in which they are performed. There are keyboard shortcuts. Each element of in this menu refers to subsequent stages described in the following Subsections. Also, each stage can save its partial results to a file, and restore at it later at need. The files of the package, unless stated otherwise, use the file extension \texttt{.edq}.

\subsection{Create Data Config Screen}

Since the Python package is tightly integrated with analyzing data from experiments, one of the first steps to use it is configuring how the experimental data is stored in files. This is defined via the so-called "Data Config"---not to be confused with the configuration of the Bell scenario, which describes how many settings and measurements each party has.

We assume that the data is entered line by line according to a specified format, and the Data Config determines which columns in each line correspond to which data. Each row gives the number of counts of various types of events for a fixed time unit and corresponds to one specific setting pair for Alice and Bob. Data from all rows are summed.

Selecting "Create Data Config..." from the "File" menu opens a dialog where you specify the number of outcomes per measurement for Alice and Bob, denoted as \texttt{A\_config} and \texttt{B\_config}, respectively. Leave the default option "2,2", meaning both parties have two binary measurements.

\begin{figure}
    \centering
    \includegraphics[width=\linewidth]{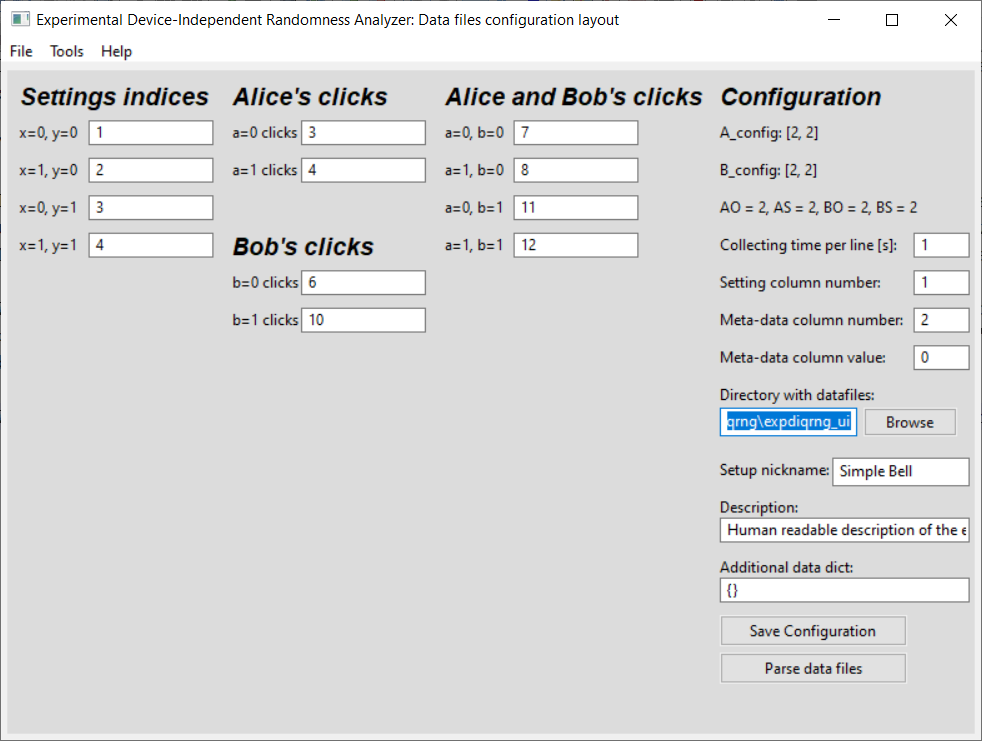}
    \caption{The Data Config screen where column mappings for experimental data are defined.}
    \label{fig:create_data_config}
\end{figure}

A screen will then appear to define the Data Config, as shown in Fig.~\ref{fig:create_data_config}.

\textbf{Settings indices} assigns an index to each pair of settings. The number of setting pairs is equal to the product of Alice's and Bob's settings. In our example, this is \(2 \times 2 = 4\). Indices 1, 2, 3, and 4 are assigned by default but can be changed.

\textbf{Alice's clicks} section defines which columns record the counts for Alice's detector events. For example, "a=0 clicks" are in column 3 (columns start from 1). Similarly, \textbf{Bob's clicks} specifies the corresponding columns for Bob. Here value of \texttt{a} and \texttt{b} label the measurement results.

\textbf{Alice and Bob's clicks} assigns columns for coincidence counts. Columns listed in \textbf{Alice's clicks} and \textbf{Bob's clicks} should include all relevant clicks, whether or not coincidences occurred.

The \textbf{Configuration} section summarizes the previous input. The values AO, AS denote the number of outcomes and settings for Alice; BO and BS for Bob. AO and BO are the maximum number of outcomes among the parties' measurements, i.e., \texttt{AO=max(A\_config)}, \texttt{BO=max(B\_config)}.

\textbf{Collecting time per line [s]} sets the time duration that each data line aggregates.

\textbf{Setting column number} specifies which column identifies the setting pair, according to the values specified in \textbf{Settings indices}.

\textbf{Meta-data column number} sets the column with metadata (e.g., experiment type). \textbf{Meta-data column value} specifies the required value; rows with a different value will be ignored.

\textbf{Directory with datafiles} sets the folder containing experimental data files (with \texttt{.dat} extension). All such files will be loaded.

\textbf{Setup nickname} and \textbf{Description} allow user-friendly descriptions.

You can save the Data Config using the "Save Configuration" button. Clicking "Parse data files" will process all \texttt{.dat} files in the selected directory using the specified Data Config. The configuration is saved as a JSON file. An example is shown in Listing~\ref{listing:example_data_config}.

\begin{lstlisting}[caption={Example Data Config JSON file.},label={listing:example_data_config},language=Python,float,firstnumber=1]
{"A_config": [2, 2], "B_config": [2, 2], "AO": 2, "BO": 2, "AS": 2, "BS": 2,
"settings_indices": [[1, 2], [3, 4]], "alice_clicks_column": [3, 4],
"bob_clicks_column": [6, 10], "alice_bob_clicks_column": [[7, 8], [11, 12]],
"time_per_line": 1.0, "setting_column_number": 1, "meta_data_column_number": 2,
"meta_data_column_value": 0, "directory_with_datafiles": "C:\\Example",
"setup_nickname": "Simple Bell", "human_description": "Human readable description 
of the experiment from datafiles", "additional_data_dict": {}}
\end{lstlisting}

A sample data file consistent with this config is shown in Listing~\ref{listing:example_data_file}. We see that the rows correspond to settings \((x, y) = (0,0), (1,0), (0,1), (1,1)\), and again (1,0). Row 5 has meta-data value 1, which is different from the desired 0, and is therefore ignored.

\begin{lstlisting}[caption={Example experimental data file.},label={listing:example_data_file},language=Python,float,firstnumber=1]
1 0 500000000 500000000 0 500000000 426776695 73223304 0 500000000 73223304 426776695
2 0 500000000 500000000 0 500000000 426776695 73223304 0 500000000 73223304 426776695
3 0 500000000 500000000 0 500000000 426776695 73223304 0 500000000 73223304 426776695
4 0 500000000 500000000 0 500000000 73223304 426776695 0 500000000 426776695 73223304
2 1 1 2 3 4 5 6 7 8 9 0
\end{lstlisting}

\subsection{Eber\_data Screen}

\begin{figure}
    \centering
    \includegraphics[width=\linewidth]{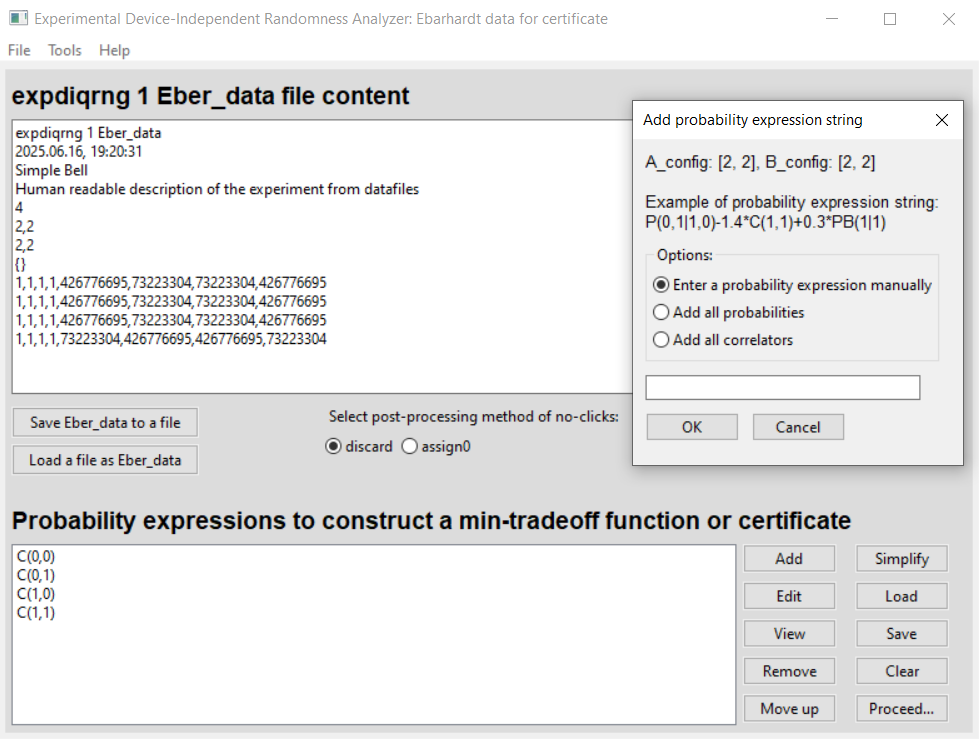}
    \caption{Screen for managing loaded experimental data (\texttt{Eber\_data}).}
    \label{fig:eber_data_screen}
\end{figure}

The next functional screen handles the so-called \texttt{Eber\_data}, which contains the parsed experimental data. See Fig.~\ref{fig:eber_data_screen}.

Click "Add" to choose which linear expressions from Alice and Bob’s conditional probabilities are included in the certificate. For instance, instead of the CHSH expression \(C(0,0) + C(0,1) + C(1,0) - C(1,1)\), we might choose to include all correlators individually: \(C(0,0)\), \(C(0,1)\), \(C(1,0)\), and \(C(1,1)\), which may yield more randomness~\cite{nieto2014using,bancal2014more}.

Click "Proceed..." to move to the next step.

\subsection{Create Certificate Screen}

\begin{figure}
    \centering
    \includegraphics[width=\linewidth]{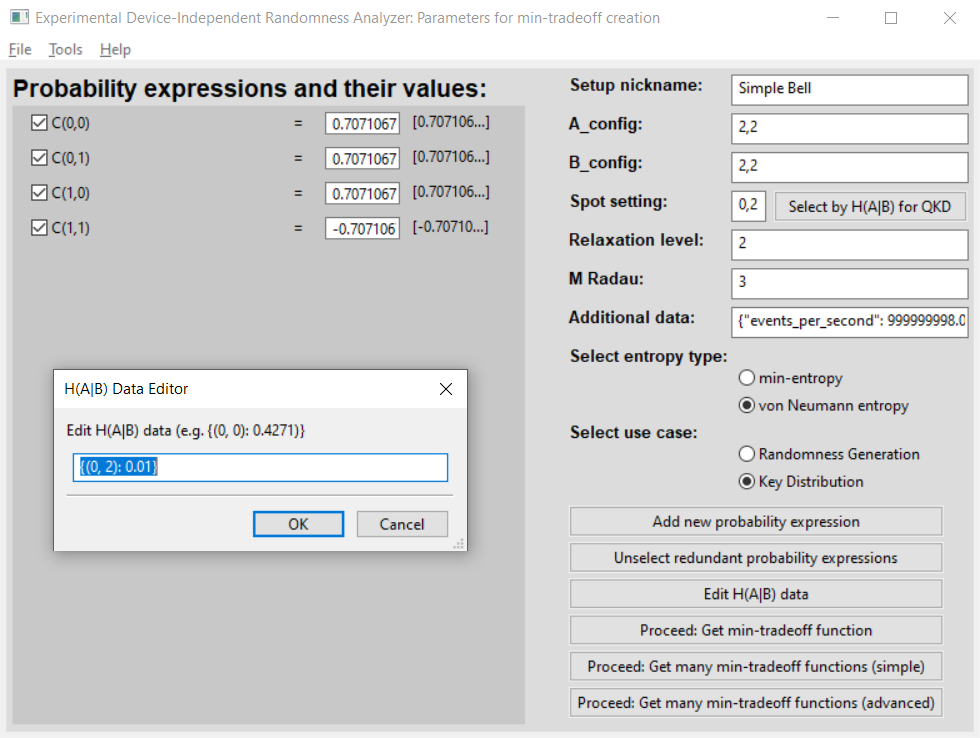}
    \caption{Create certificate screen with options for entropy type and spot-setting.}
    \label{fig:create_certificate_screen}
\end{figure}

The next screen, shown in Fig.~\ref{fig:create_certificate_screen}, lets the user create the certificate used to derive the min-tradeoff function. In addition to the expressions computed from experimental data, the user can add fixed linear expressions.

We now aim to perform QKD analysis. For this, we need the error correction term from the Devetak–Winter formula~\cite{devetak2005distillation}:
\begin{equation}
	r = H(A|E) - H(A|B).
\end{equation}
Here, \(H(A|B)\) is the error correction term for the chosen spot-setting~\cite{miller2017universal}. In spot-checking QRNG protocols, one pair of settings is used during the key generation phase, and the rest are used for test rounds.

Choose entropy type as \texttt{von~Neumann entropy}, and use case as \texttt{Key Distribution}.

Choose the spot-setting \((0,2)\). Note that outcome 2 for Bob wasn’t included in the earlier data, but the tool optionally allows specifying it here. Click "Edit H(A|B) data" and enter \verb!{(0, 2): 0.01}!, meaning the error correction term is 0.01 for that setting. Now click "Proceed: Get min-tradeoff function" to compute the min-tradeoff function.

\subsection{Rate Calculation and Presentation}

\begin{figure}
    \centering
    \includegraphics[width=\linewidth]{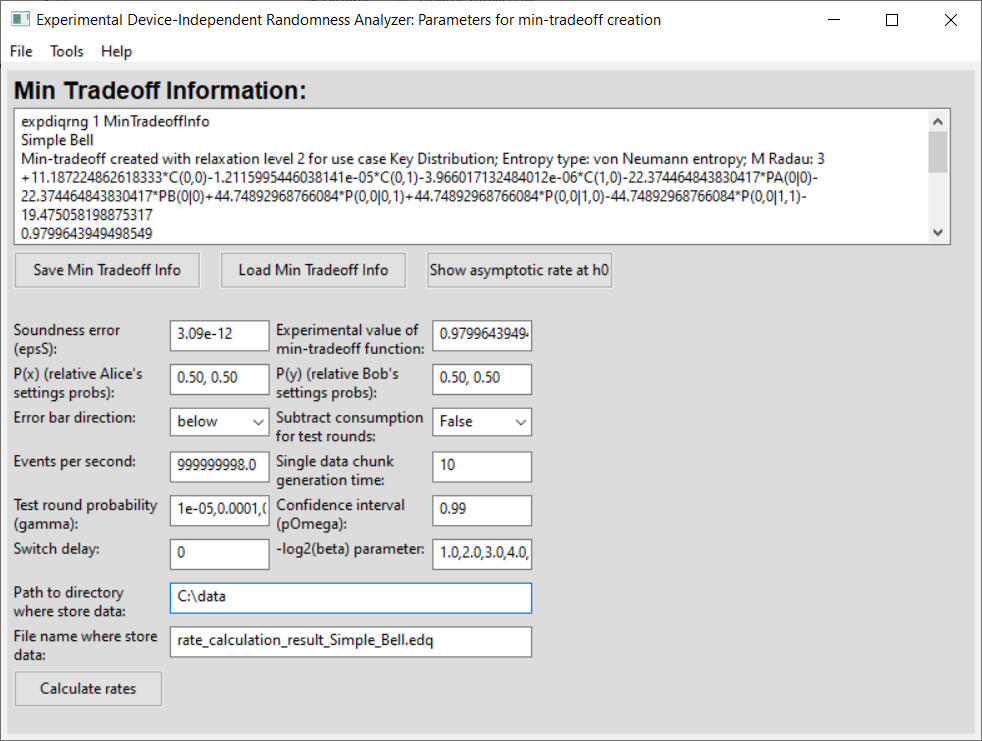}
    \caption{Screen for computing and displaying the min-tradeoff function.}
    \label{fig:min_tradeoff_screen}
\end{figure}

Fig.~\ref{fig:min_tradeoff_screen} shows the computed min-tradeoff function. For this example, use all default values. The sample data from Listing~\ref{listing:example_data_file} has \(10^9\) events per second. Set "Single data chunk generation time" to 10 seconds and click "Calculate rates".

\begin{figure}
    \centering
    \includegraphics[width=\linewidth]{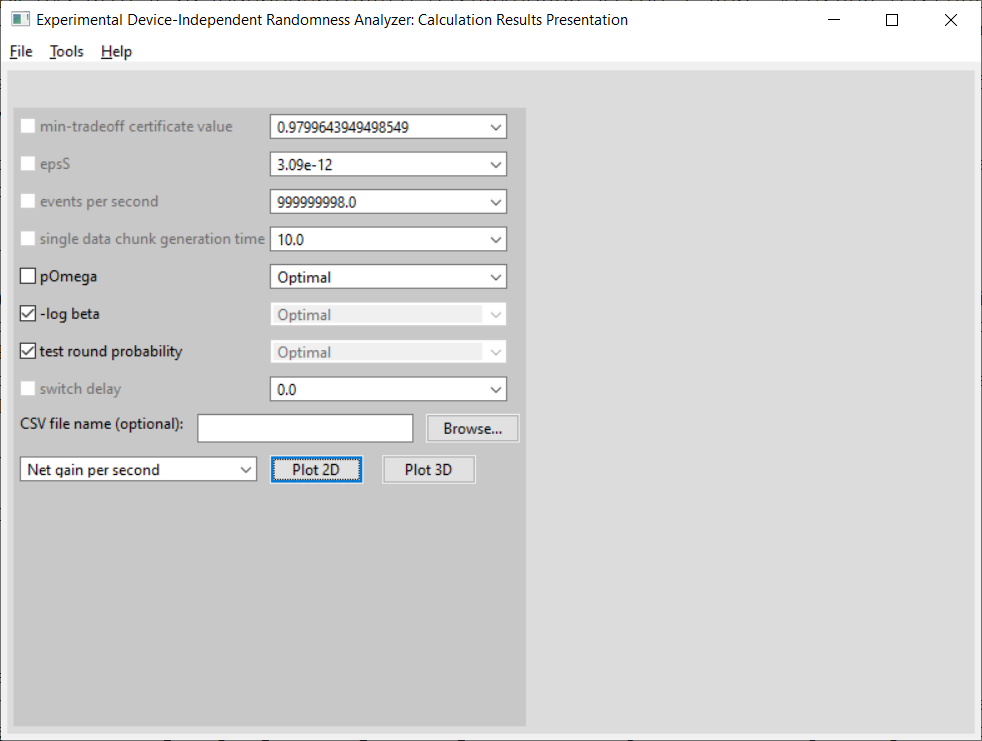}
    \caption{Results screen to visualize key length vs. parameters.}
    \label{fig:results_presentation_screen}
\end{figure}

Fig.~\ref{fig:results_presentation_screen} illustrates plotting how the key length depends on parameters from Fig.~\ref{fig:min_tradeoff_screen}. Select "-log beta" and "test round probability", then click "Plot 2D".

\begin{figure}
    \centering
    \includegraphics[width=\linewidth]{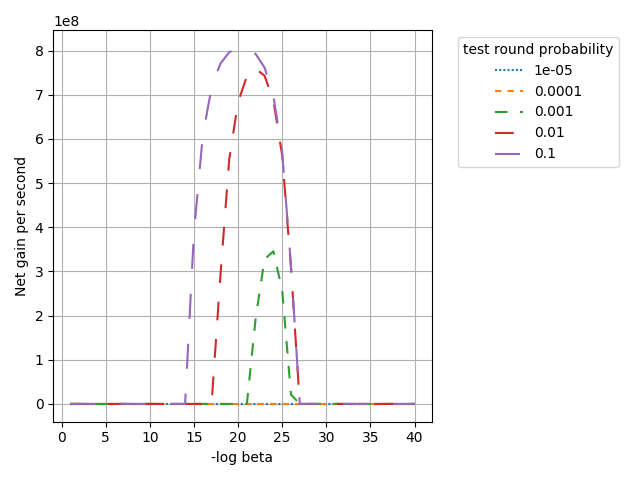}
    \caption{Generated key bits per second vs. test round probability \(\gamma\) and parameter \(\beta\).}
    \label{fig:gain_gamma_beta}
\end{figure}

From Fig.~\ref{fig:gain_gamma_beta}, for test round probability \(\gamma = 0.1\), we obtain about \(8 \times 10^8\) bits/second. The min-tradeoff certificate value is approximately 0.979964, and error correction term \(H(A|B)\) was 0.01, so the asymptotic keyrate per event is roughly 0.96. Given nearly \(10^9\) events/second, this gives a maximum asymptotic keyrate near \(9.6 \times 10^8\) bits/second. The plot shows that for \(\beta \approx 2^{-20}\), we achieve the aforementioned \(8 \times 10^8\) bits/second. The difference is due to finite-size effects accounted for by EAT.

\section{Conclusions}\label{sec:conclusions}

We have presented a comprehensive framework for automating and analyzing quantum cryptographic protocols based on the Entropy Accumulation Theorem (EAT). Our software integrates both command-line API and a graphical user interface, enabling flexible use for a variety of experimental setups and entropy measures, including min-entropy and von~Neumann entropy.

Through two detailed case studies—QRNG in a modified CHSH scenario and QKD in a standard CHSH scenario—we demonstrated the practicality of our tool for end-to-end analysis, from parsing raw data to computing finite-size secure key rates. The software allows for customization of Bell expressions, entropy types, and protocol parameters, making it a valuable resource for experimental quantum cryptography.

Future work may include extending support for more general Bell scenarios (e.g. including more parties), automating semi-definite program selection, additional constraints (e.g. dimension~\cite{mironowicz2014properties,navascues2015bounding}), parallel repetitions~\cite{miklin2022exponentially}, wirings~\cite{forster2009distilling,ulu2025device}, further optimizing computational performance for large-scale data sets~\cite{mironowicz2018applications}, and explicit randomness extraction features~\cite{de2012trevisan,mauerer2012modular,ma2013postprocessing}.

\section*{Acknowledgements}
This work was supported by the Knut and Alice Wallenberg Foundation through the Wallenberg Centre for Quantum Technology (WACQT). PM was supported by the European Union’s Horizon Europe research and innovation programme under grant agreement No 101080086/NeQST.

\section*{Author contributions}
PM developed the \texttt{expdiqrng} and \texttt{expdiqrng\_ui} software packages, implementing both the front-end and back-end. MB proposed the idea of building a graphical toolbox for finite-size device-independent analysis using the Entropy Accumulation Theorem. MB also provided feedback and guidance to align the software with experimental needs and data formats used in his laboratory. Both authors contributed to discussions on the functionality of the software.

\section*{Data availability}
Please note that upon the upload of the first arXiv draft of this paper, an initial version (v0.1) of the software packages was released. The code will be maintained and updated based on community feedback, including the addition of new features and the implementation of bug fixes. Additionally, the current version already includes several advanced features not discussed in detail above.
The source code of the \texttt{expdiqrng} and \texttt{expdiqrng\_ui} packages, including the graphical interface and examples for randomness generation and key distribution, is publicly available on GitHub at \url{https://github.com/PiotrMironowicz/expdiqrng}. All data required to reproduce the examples and figures in this manuscript are included in the repository.

\appendix

\section{Semi-definite programming and Brown–Fawzi–Fawzi relaxations}\label{sec:BFF}

Semi-definite programming (SDP) is a powerful optimization framework designed to address problems involving linear constraints and objective, where the optimization variables are matrix-valued and constrained to be positive semidefinite. A matrix is positive semidefinite if it is Hermitian—equal to its own conjugate transpose—and has non-negative eigenvalues~\cite{mironowicz2024semi}.

SDP can be used to provide lower-bound on the von~Neumann entropy using the Brown-Fawzi-Fawzi (BFF) method~\cite{brown2024device}, described below. Consider the tripartite Hilbert space \( Q_A \otimes Q_B \otimes Q_E \), associated with Alice, Bob, and an adversary Eve, respectively, and let \( \rho_{Q_A Q_B Q_E} \) denote their shared quantum state. Our goal is to numerically compute a rigorous lower bound on the conditional von Neumann entropy, expressed as:
\begin{equation}
	\label{eq:Habxye}
	H(Q_A Q_B \mid x = x^*, y = y^*, Q_E)_{\rho_{Q_A Q_B Q_E}}.
\end{equation}
Here, the families \( \{ \{ M_{a|x} \}_a \}_x \) and \( \{ \{ N_{b|y} \}_b \}_y \) represent the positive operator-valued measurements (POVMs) employed by Alice and Bob, respectively. Utilizing the Gauss–Radau quadrature method, a lower bound on the conditional entropy in \eqref{eq:Habxye} can be obtained:
\begin{equation}
	\label{eq:opt}
	\sum_{i} c_i \sum_{a,b = 0,1} \inf_{\substack{Z_{a,b} \in B(Q_E), \ \mathrm{cond}(P)}} \left( 1 + F[M_{a|x^*}, N_{b|y^*}, Z_{a,b}, t_i] \right),
\end{equation}
where the function \( F[M_{a|x^*}, N_{b|y^*}, Z_{a,b}, t_i] \) is given by the trace
\begin{equation}
	F = \Tr \left[ \rho_{Q_A Q_B Q_E} \left( O_1 + O_2 \right) \right].
\end{equation}
The constraint \( \mathrm{cond}(P) \) enforces compatibility with the observed distribution
\begin{equation}
	P(a,b \mid x, y) \equiv \Tr\left[ \rho_{Q_A Q_B} \, M_{a|x^*} \otimes N_{b|y^*} \right],
\end{equation}
which must satisfy linear conditions specified by the quantum protocol. The operators \( O_1 \) and \( O_2 \) are defined as:
\begin{subequations}
	\begin{equation}
		O_1 \equiv M_{a|x^*} \otimes N_{b|y^*} \otimes \left( Z_{a,b} + Z_{a,b}^\dagger + (1 - t_i) Z_{a,b} Z_{a,b}^\dagger \right),
	\end{equation}
	\begin{equation}
		O_2 \equiv t_i \left( \openone_{Q_A Q_B} \otimes Z_{a,b} Z_{a,b}^\dagger \right).
	\end{equation}
\end{subequations}
The weights \( c_i \) are derived from the Gauss–Radau quadrature formula as
\begin{equation}
	c_i \equiv \frac{w_i}{t_i \log(2)},
\end{equation}
where \( w_i \) and \( t_i \) are the weights and nodes of the quadrature rule, respectively. The index \( i \) runs over all quadrature nodes except the final one.

This semidefinite relaxation enables us to obtain certified lower bounds on the entropy that are both numerically tractable and information-theoretically valid in quantum adversarial scenarios.


\begin{thebibliography}{10}

\bibitem{gisin2002quantum}
N.~Gisin, G.~Ribordy, W.~Tittel, and H.~Zbinden, ``Quantum cryptography,'' {\em
  Reviews of modern physics}, vol.~74, no.~1, p.~145, 2002.

\bibitem{scarani2009security}
V.~Scarani {\em et~al.}, ``The security of practical quantum key
  distribution,'' {\em Reviews of Modern Physics}, vol.~81, no.~3,
  pp.~1301--1350, 2009.

\bibitem{nielsen2010quantum}
M.~A. Nielsen and I.~L. Chuang, {\em Quantum Computation and Quantum
  Information}.
\newblock Cambridge University Press, 2010.

\bibitem{pirandola2020advances}
S.~Pirandola, U.~L. Andersen, L.~Banchi, M.~Berta, D.~Bunandar, R.~Colbeck,
  D.~Englund, T.~Gehring, C.~Lupo, C.~Ottaviani, {\em et~al.}, ``Advances in
  quantum cryptography,'' {\em Advances in optics and photonics}, vol.~12,
  no.~4, pp.~1012--1236, 2020.

\bibitem{pironio2010random}
S.~Pironio {\em et~al.}, ``Random numbers certified by {B}ell's theorem,'' {\em
  Nature}, vol.~464, pp.~1021--1024, 2010.

\bibitem{acin2016certified}
A.~Ac{\'\i}n and L.~Masanes, ``Certified randomness in quantum physics,'' {\em
  Nature}, vol.~540, pp.~213--219, 2016.

\bibitem{tomamichel2016quantum}
M.~Tomamichel, {\em Quantum Information Processing with Finite Resources:
  Mathematical Foundations}, vol.~5 of {\em SpringerBriefs in Mathematical
  Physics}.
\newblock Springer, Cham, 1~ed., 2016.
\newblock 4 illustrations in colour.

\bibitem{dupuis2020entropy}
F.~Dupuis, O.~Fawzi, and R.~Renner, ``Entropy accumulation,'' {\em
  Communications in Mathematical Physics}, vol.~379, pp.~867--913, 2020.

\bibitem{knill2018quantum}
E.~Knill, Y.~Zhang, and H.~Fu, ``Quantum probability estimation for randomness
  with quantum side information,'' {\em arXiv preprint arXiv:1806.04553}, 2018.

\bibitem{clauser1969proposed}
J.~F. Clauser, M.~A. Horne, A.~Shimony, and R.~A. Holt, ``Proposed experiment
  to test local hidden-variable theories,'' {\em Physical Review Letters},
  vol.~23, no.~15, p.~880, 1969.

\bibitem{brown2019framework}
P.~J. Brown, S.~Ragy, and R.~Colbeck, ``A framework for quantum-secure
  device-independent randomness expansion,'' {\em IEEE Transactions on
  Information Theory}, 2019.

\bibitem{dupuis2019entropy}
F.~Dupuis and O.~Fawzi, ``Entropy accumulation with improved second-order
  term,'' {\em IEEE Transactions on Information Theory}, 2021.
\newblock to appear.

\bibitem{metger2024generalised}
T.~Metger, O.~Fawzi, D.~Sutter, and R.~Renner, ``Generalised entropy
  accumulation,'' {\em Communications in Mathematical Physics}, vol.~405,
  no.~11, p.~261, 2024.

\bibitem{arqand2024generalized}
A.~Arqand, T.~A. Hahn, and E.~Y.-Z. Tan, ``Generalized r$\backslash$'enyi
  entropy accumulation theorem and generalized quantum probability
  estimation,'' {\em arXiv preprint arXiv:2405.05912}, 2024.

\bibitem{arnon2018practical}
R.~Arnon-Friedman, F.~Dupuis, O.~Fawzi, R.~Renner, and T.~Vidick, ``Practical
  device-independent quantum cryptography via entropy accumulation,'' {\em
  Nature communications}, vol.~9, no.~1, p.~459, 2018.

\bibitem{metger2023security}
T.~Metger and R.~Renner, ``Security of quantum key distribution from
  generalised entropy accumulation,'' {\em Nature Communications}, vol.~14,
  no.~1, p.~5272, 2023.

\bibitem{arnon2019simple}
R.~Arnon-Friedman, R.~Renner, and T.~Vidick, ``Simple and tight
  device-independent security proofs,'' {\em SIAM Journal on Computing},
  vol.~48, no.~1, pp.~181--225, 2019.

\bibitem{liu2021device}
W.-Z. Liu, M.-H. Li, S.~Ragy, S.-R. Zhao, B.~Bai, Y.~Liu, P.~J. Brown,
  J.~Zhang, R.~Colbeck, J.~Fan, {\em et~al.}, ``Device-independent randomness
  expansion against quantum side information,'' {\em Nature Physics}, vol.~17,
  no.~4, pp.~448--451, 2021.

\bibitem{nadlinger2022experimental}
D.~P. Nadlinger, P.~Drmota, B.~C. Nichol, G.~Araneda, D.~Main, R.~Srinivas,
  D.~M. Lucas, C.~J. Ballance, K.~Ivanov, E.-Z. Tan, {\em et~al.},
  ``Experimental quantum key distribution certified by {B}ell's theorem,'' {\em
  Nature}, vol.~607, no.~7920, pp.~682--686, 2022.

\bibitem{brunner2014bell}
N.~Brunner, D.~Cavalcanti, S.~Pironio, V.~Scarani, and S.~Wehner, ``Bell
  nonlocality,'' {\em Reviews of modern physics}, vol.~86, no.~2, pp.~419--478,
  2014.

\bibitem{miller2017universal}
C.~A. Miller and Y.~Shi, ``Universal security for randomness expansion from the
  spot-checking protocol,'' {\em SIAM Journal on Computing}, vol.~46, no.~4,
  pp.~1304--1335, 2017.

\bibitem{piveteau2024optimization}
A.~Piveteau, A.~Seguinard, P.~Mironowicz, and M.~Bourennane, ``Optimization of
  experimental quantum randomness expansion,'' {\em arXiv preprint
  arXiv:2411.04934}, 2024.

\bibitem{navascues2007bounding}
M.~Navascu{\'e}s, S.~Pironio, and A.~Ac{\'\i}n, ``Bounding the set of quantum
  correlations,'' {\em Physical Review Letters}, vol.~98, no.~1, p.~010401,
  2007.

\bibitem{navascues2008convergent}
M.~Navascu{\'e}s, S.~Pironio, and A.~Ac{\'\i}n, ``A convergent hierarchy of
  semidefinite programs characterizing the set of quantum correlations,'' {\em
  New Journal of Physics}, vol.~10, no.~7, p.~073013, 2008.

\bibitem{mironowicz2024semi}
P.~Mironowicz, ``Semi-definite programming and quantum information,'' {\em
  Journal of Physics A: Mathematical and Theoretical}, vol.~57, no.~16,
  p.~163002, 2024.

\bibitem{brown2024device}
P.~Brown, H.~Fawzi, and O.~Fawzi, ``Device-independent lower bounds on the
  conditional von neumann entropy,'' {\em Quantum}, vol.~8, p.~1445, 2024.

\bibitem{nieto2014using}
O.~Nieto-Silleras, S.~Pironio, and J.~Silman, ``Using complete measurement
  statistics for optimal device-independent randomness evaluation,'' {\em New
  Journal of Physics}, vol.~16, no.~1, p.~013035, 2014.

\bibitem{bancal2014more}
J.-D. Bancal, L.~Sheridan, and V.~Scarani, ``More randomness from the same
  data,'' {\em New Journal of Physics}, vol.~16, no.~3, p.~033011, 2014.

\bibitem{wittek2015algorithm}
P.~Wittek, ``Algorithm 950: Ncpol2sdpa—sparse semidefinite programming
  relaxations for polynomial optimization problems of noncommuting variables,''
  {\em ACM Transactions on Mathematical Software (TOMS)}, vol.~41, no.~3,
  pp.~1--12, 2015.

\bibitem{Ncpol2sdpaGitHub}
P.~Wittek and P.~J. Brown, ``Updating ncpol2sdpa after {Peter Wittek}.''
  \url{https://github.com/peterjbrown519/ncpol2sdpa}.

\bibitem{mironowicz2013robustness}
P.~Mironowicz and M.~Paw{\l}owski, ``Robustness of quantum-randomness expansion
  protocols in the presence of noise,'' {\em Physical Review A—Atomic,
  Molecular, and Optical Physics}, vol.~88, no.~3, p.~032319, 2013.

\bibitem{acin2006efficient}
A.~Acin, S.~Massar, and S.~Pironio, ``Efficient quantum key distribution secure
  against no-signalling eavesdroppers,'' {\em New Journal of Physics}, vol.~8,
  no.~8, p.~126, 2006.

\bibitem{devetak2005distillation}
I.~Devetak and A.~Winter, ``Distillation of secret key and entanglement from
  quantum states,'' {\em Proceedings of the Royal Society A: Mathematical,
  Physical and engineering sciences}, vol.~461, no.~2053, pp.~207--235, 2005.

\bibitem{mironowicz2014properties}
P.~Mironowicz, H.-W. Li, and M.~Paw{\l}owski, ``Properties of dimension
  witnesses and their semidefinite programming relaxations,'' {\em Physical
  Review A}, vol.~90, no.~2, p.~022322, 2014.

\bibitem{navascues2015bounding}
M.~Navascu{\'e}s and T.~V{\'e}rtesi, ``Bounding the set of finite dimensional
  quantum correlations,'' {\em Physical Review Letters}, vol.~115, no.~2,
  p.~020501, 2015.

\bibitem{miklin2022exponentially}
N.~Miklin, A.~Chaturvedi, M.~Bourennane, M.~Paw{\l}owski, and A.~Cabello,
  ``Exponentially decreasing critical detection efficiency for any {B}ell
  inequality,'' {\em Physical Review Letters}, vol.~129, no.~23, p.~230403,
  2022.

\bibitem{forster2009distilling}
M.~Forster, S.~Winkler, and S.~Wolf, ``Distilling nonlocality,'' {\em Physical
  Review Letters}, vol.~102, no.~12, p.~120401, 2009.

\bibitem{ulu2025device}
B.~Ulu, N.~Brunner, and M.~Weilenmann, ``Device independent quantum key
  activation,'' {\em arXiv preprint arXiv:2506.09772}, 2025.

\bibitem{mironowicz2018applications}
P.~Mironowicz, ``Applications of semi-definite optimization in quantum
  information protocols,'' {\em arXiv preprint arXiv:1810.05145}, 2018.

\bibitem{de2012trevisan}
A.~De, C.~Portmann, T.~Vidick, and R.~Renner, ``Trevisan's extractor in the
  presence of quantum side information,'' {\em SIAM Journal on Computing},
  vol.~41, no.~4, pp.~915--940, 2012.

\bibitem{mauerer2012modular}
W.~Mauerer, C.~Portmann, and V.~B. Scholz, ``A modular framework for randomness
  extraction based on trevisan's construction,'' {\em arXiv preprint
  arXiv:1212.0520}, 2012.

\bibitem{ma2013postprocessing}
X.~Ma, F.~Xu, H.~Xu, X.~Tan, B.~Qi, and H.-K. Lo, ``Postprocessing for quantum
  random-number generators: Entropy evaluation and randomness extraction,''
  {\em Physical Review A—Atomic, Molecular, and Optical Physics}, vol.~87,
  no.~6, p.~062327, 2013.

\end{thebibliography}
\end{document}